\documentclass[aps,prl,floatfix,twocolumn,showpacs]{revtex4}
\usepackage{amssymb}
\usepackage{amsmath}
\usepackage{graphicx}

\begin{document}

\title{Shot noise in carbon nanotube based Fabry-Perot interferometers}
\author{L.G. Herrmann$^{1}$, T. Delattre$^{1}$, P. Morfin$^{1}$, J.-M. Berroir$^{1}$, B. Pla\c cais$^{1}$, D.C. Glattli$^{1,2}$ and T. Kontos$^{1}$}
\affiliation{$^{1}$Laboratoire Pierre Aigrain, Ecole Normale Sup\'erieure, 24, rue Lhomond, 75231 Paris Cedex 05, France\\
$^{2}$Service de physique de l'\'etat Condens\'e, CEA, 91192
Gif-sur-Yvette, France.}
\pacs{73.23.-b,72.70.+m,73.63.Fg}

\begin{abstract}
We report on shot noise measurements in carbon nanotube based
Fabry-Perot electronic interferometers. As a consequence of quantum
interferences, the noise power spectral density oscillates as a
function of the voltage applied to the gate electrode. The quantum
shot noise theory accounts for the data quantitatively. It allows to
confirm the existence of two nearly degenerate orbitals. At
resonance, the transmission of the nanotube approaches unity, and
the nanotube becomes noiseless, as observed in quantum point
contacts. In this weak backscattering regime, the dependence of the
noise on the backscattering current is found weaker than expected,
pointing either to electron-electron interactions or to weak
decoherence.
\end{abstract}

\date{\today}
\maketitle

The quantum character of transport in mesoscopic conductors
qualitatively modifies the behavior of both the average and the
fluctuations of the current that flows through them
\cite{Blanter:00}. If interactions between charge carriers can be
neglected, an accurate description of such conductors is given by a
set of transmission probabilities $\{T_n\}$ which characterize the
scattering of carriers. This description has been tested
successfully for current noise in various conductors, ranging from
quantum point contacts (QPCs) \cite{Kumar:96}, in which one can
isolate one spin degenerate channel with a single tunable barrier,
to superconducting/normal/superconducting (S/N/S) structures
\cite{Jehl:00,Cron:01,Hoss:00}. In coherent few channel double
barrier systems, quantum interference have also been shown to
modulate the transmissions \cite{vanWees:89}. However, shot noise in
such Fabry-Perot electronic interferometers has not been
investigated experimentally so far.

\begin{figure}[!pth]
\centering\includegraphics[height=0.5\linewidth,angle=0]{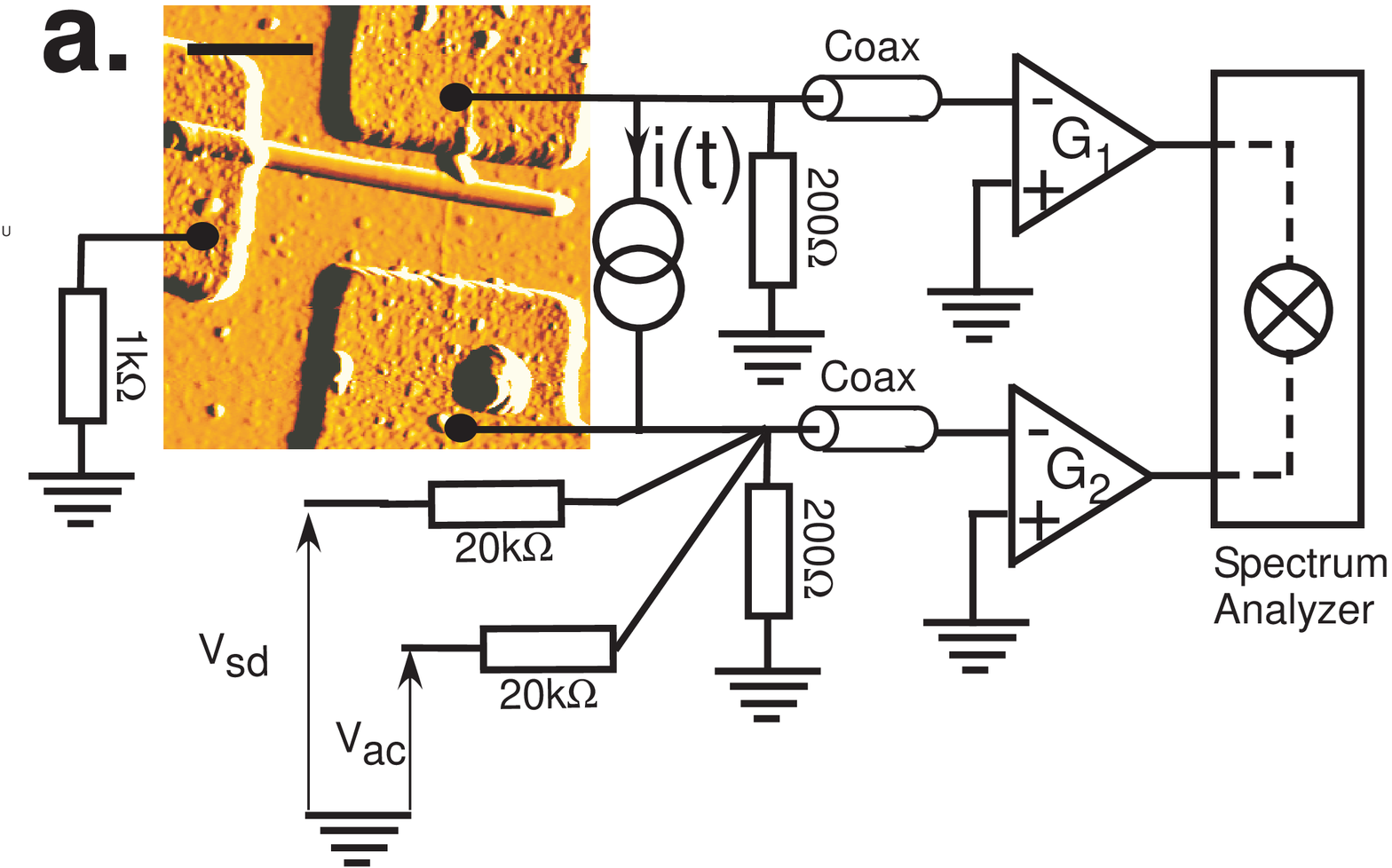}
\quad\includegraphics[height=0.55\linewidth,angle=0]{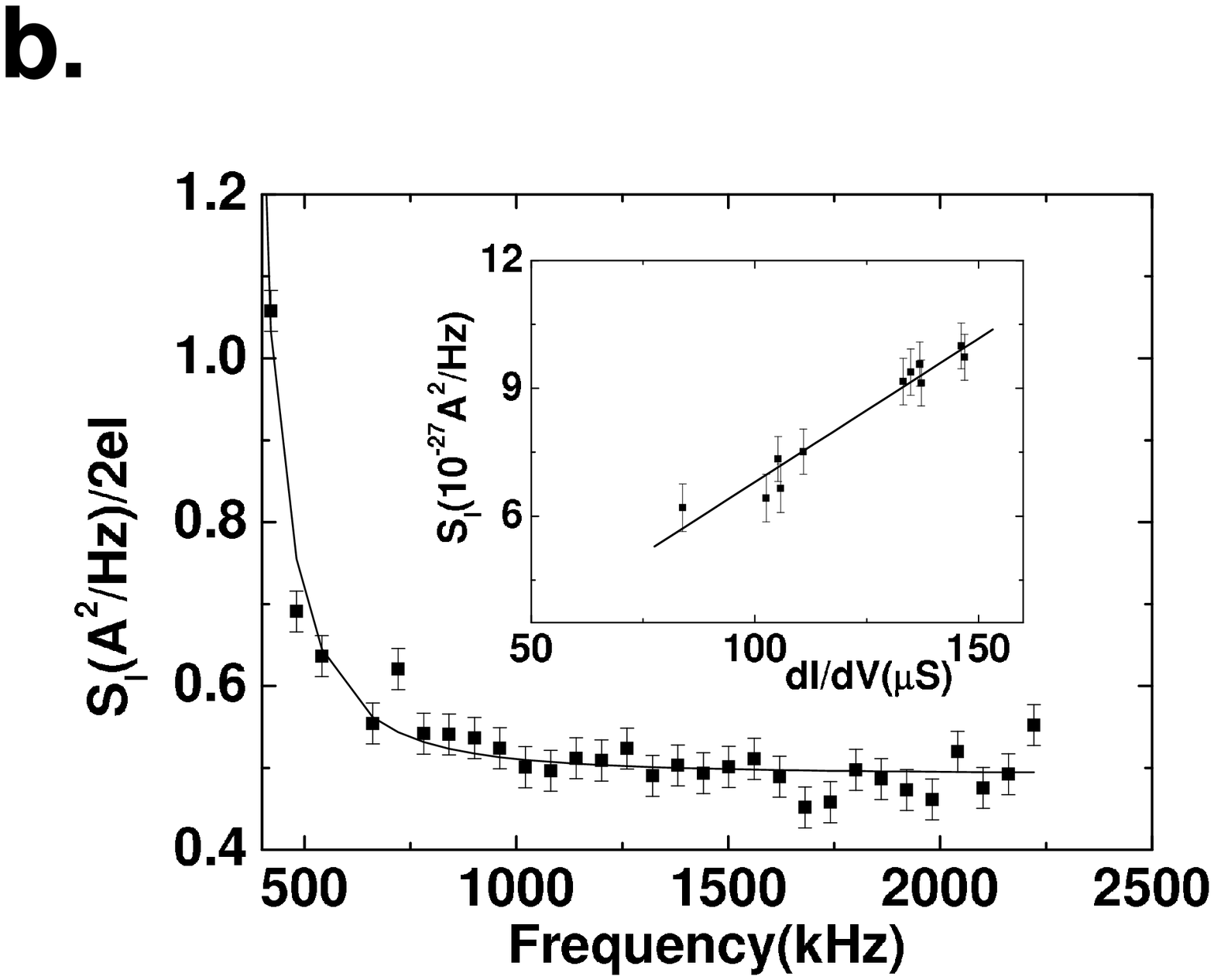}
\caption{a. Diagram of the circuit and AFM picture of the main
sample presented in this paper. The MWNT is shunted by a $1k\Omega$
resistor placed on the PCB. The bar is $500nm$. b. Frequency
dependence of the noise power spectral density measured at
$V_{SD}=1.1 mV$ normalized by the Schottky value $2eI_{SD}$.
The solid line is a fit with the formula
$0.49+92.6/((f-300)^2+(50)^2)$. Inset: Calibration of the background
noise as a function of the conductance of the NT.
}%
\label{diagramm}%
\end{figure}

Single Wall carbon NanoTubes (SWNTs) can display a Fabry-Perot
behavior \cite{bockrath:01,Dai:05} when their coupling to metallic
reservoirs is high enough. Due to the so-called K-K' (orbital)
degeneracy, two transmissions $\{T_{1},T_{2}\}$ are needed in
general to characterize transport in these
devices\cite{bockrath:01}. Therefore, the combined measurement of
noise and conductance should allow a full characterization of any
nanotube  in this regime by determination of this set. Early
measurements of current noise in carbon nanotubes have shown that it
was dominated by extrinsic $1/f$ noise below $100kHz$
\cite{noise1of:99}. For this reason, few shot noise measurements are
available. In ref. \cite{Onac:06}, the Coulomb blockade regime was
investigated and significant departures from the predictions of the
non-interacting theory were found. In ref. \cite{Reulet:99}, very
low shot noise was found in a bundle of SWNTs highly coupled to
normal reservoirs, pointing to ballistic transport. Very recently,
the high bias shot noise has been investigated in gated carbon
nanotube based Fabry-Perot interferometers \cite{Yamamoto:06} and
signatures of electron-electron interactions have been found.

In this letter, we report on shot noise measurements in gated SWNTs
in the low energy Fabry-Perot regime \cite{Pertti:07}. This allows a
reliable quantitative comparison with the quantum shot noise theory.
The measurement frequency, ranging from $400kHz$ to $5MHz$, makes
the intrinsic (shot) noise dominate. Cross-correlations techniques
\cite{glattli:97} and room-temperature ultra-low noise
preamplification give the required sensitivity (see EPAPS for
details). We find that the nanotube is well described by a
non-interacting scattering theory accounting for the so-called K-K'
orbital degeneracy commonly found in NTs and arising from the
band-structure of graphene\cite{KK}. For the sample presented in
this letter, the transmissions for the two channels are found to be
equal within $10\%$. Near Fabry-Perot resonances, for which the
transmission is close to $1$, shot noise is strongly suppressed, as
expected. However, its dependence with the backscattering current is
found weaker than expected. This might be due to electron-electron
interactions or weak decoherence.

The SWNTs are grown by chemical vapor deposition with a standard
recipe \cite{Dai:05}. They are localized with respect to alignment
markers with an atomic force microscope (AFM). The contacts are made
by e-beam lithography followed by evaporation of a $70nm$-thick Pd
layer at a pressure of $10^{-8} mbar$. The highly doped Si substrate
covered with $500nm$ doped $SiO_{2}$ is used as a back-gate at low
temperatures. The typical spacing between the Pd electrodes is
$500nm$ as shown in figure \ref{diagramm}a. The two probe resistance
of the obtained devices ranges from $10k\Omega$ to $200k\Omega$ at
room temperature. For some samples, a third probe made of a
Multi-Wall carbon NanoTube (MWNT) is placed with the help of the AFM
tip on the top of the SWNT. Although the sample presented in this
paper is of this kind, as shown in figure \ref{diagramm}a, the
coupling of the SWNT with the MWNT is very weak and can be omitted
in the diagram. The temperature is $1.5K$ unless specified.

The circuit diagram (see figure \ref{diagramm}a) yields the
relationship between the voltage correlations
$S_{cross}=<V_{1}V^{*}_{2}>$ and the different current noise sources
which contribute to the voltage fluctuations along the $200\Omega$
resistors $R_1$ and $R_2$. It turns out that the main contributions
to these fluctuations arise from the current noise $S_{I}$ in the
SWNT, the current noise of the two low noise preamplifiers
respectively $S_{n1}$ and $S_{n2}$ and the Johnson-Nyquist noise
$S_{1}$ and $S_{2}$ of $R_1$ and $R_2$, respectively . The complex
value of $S_{cross}$ is:

\begin{eqnarray}\label{eq:circuit}
S_{cross}=\mid \alpha \mid^{-2}Z_{1}Z^{*}_{2}\Big[-S_{I}+S_{off}\Big]\\
S_{off}=\frac{Z_{1}}{R_{NT}}(S_{n1}+S_{1})
+\frac{Z^{*}_{2}}{R_{NT}}(S_{n2}+S_{2})\nonumber
\end{eqnarray}

with
$\alpha=1+(Z_{1}+Z_{2})/R_{NT}+Z_{1}/R_{NT}+Z_{1}Z_{2}/(R_{NT}R_{B1})$,
where $R_{NT}$ is the resistance of the SWNT, $R_{B1}$ is the bias
resistor of line 1, $Z_{1(2)}=R_{1(2)}/(1+2\pi j R_{1(2)}C_{1(2)}
f)$, $C_{1(2)}$ the total capacitance in parallel with $R_{1(2)}$
and $f$ the frequency.
The linear behavior of the
offset noise $S_{off}$ as a function of $T$ predicted by formula
(\ref{eq:circuit}) for the sample presented in this letter is shown
in the inset of figure \ref{diagramm}b. A linear fit gives
$S_{off}=(0.05\pm 1.01 + 10.44 T\pm 1.27 T)\times 10^{-27} A^2/Hz$.
Although the order of magnitude is correct, confirming that most of
the signal comes from the noise of the NT, this is only in
qualitative agreement with the expected offset of $21.1T \times
10^{-27}A^2/Hz$. We think that this can be explained by residual
correlations arising from partial shielding of the parasitic signals
in the band $0-10MHz$.

\begin{figure}[pth]
\centering\includegraphics[height=0.65\linewidth,angle=0]{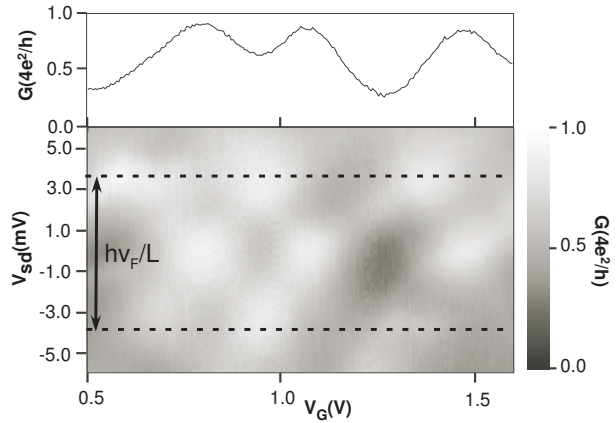}%
\caption{Linear conductance and greyscale plot of the non-linear
conductance. The characteristic checker-board pattern of a
Fabry-Perot interferometer is observed. The level spacing (black
double arrow) is of about $3.4meV$.}%
\label{frequency}%
\end{figure}

In general, the noise power spectral density displays a strong
frequency dependence which is of $1/f^{\alpha}$ type
\cite{noise1of:99,Zorin:88,Liu:06} at frequencies of the order of
$100kHz$ in nanotube devices. This extrinsic contribution points to
the effect of charge traps which are effective in Coulomb blockade
devices and could explain the observed noise in nanotubes at high
current $\sim 100nA$ and high temperature $\sim 300K$. We have
measured the noise on four different samples using the technique
depicted above. Two of them (without the MWNT) had resistances of
$250k\Omega$ and $1M\Omega$ respectively at room temperature and
exhibited Poissonian noise, as expected for a conductor with a low
transmission. For these samples, no frequency dependent noise was
observed up to the highest bias voltage applied $V_{sd}=30mV$ and
down to the lowest frequency $580kHz$ at $1.5K$. The two others
(with the MWNT) had a resistance of about $15k\Omega$ at room
temperature and $V_{G}=0V$. The resistance between the MWNT and the
SWNT was about $1M\Omega$ at room temperature going up to about
$10M\Omega$ at $1.5K$, turning these three terminal samples into
essentially two terminal ones. We will focus on one of these samples
for the remaining of the paper. Figure \ref{diagramm}b displays the
frequency dependence of the low temperature noise power spectral
density $S_I$ measured at $V_{sd}=1.25mV$ for frequencies ranging
from $421kHz$ to $2.221MHz$. This noise power spectral density has
been normalized to the Schottky value $2eI$, $I$ being the current
flowing through the device and $e$ being the elementary charge. From
$1MHz$ to $2.221MHz$, $S_I/2eI$ is roughly constant, equal to
$0.48$, up to error bars. Below $1MHz$, the noise has roughly a
$1/f^2$ dependence and the overall dependence is well fitted by the
Lorentzian line shape $0.49+92.6/((f-300)^2+(50)^2)$, with $f$ in
$kHz$. This shows that, at this bias, few fluctuators with a
characteristic frequency of about $300kHz$ are excited. Below
$V_{sd}=1mV$, at our operating frequency of $2.221MHz$, the effect
of charge fluctuators appears usually as an asymmetric noise curve
with respect to the bias. This occurs rarely, about $5\%$ of the
gate voltage range, and produces a sudden change of $S_I$ at
constant $V_{sd}$ when the gate voltage is swept. Data in this
regime are not presented for clarity.

\begin{figure}[pth]
\centering\includegraphics[height=0.65\linewidth,angle=0]{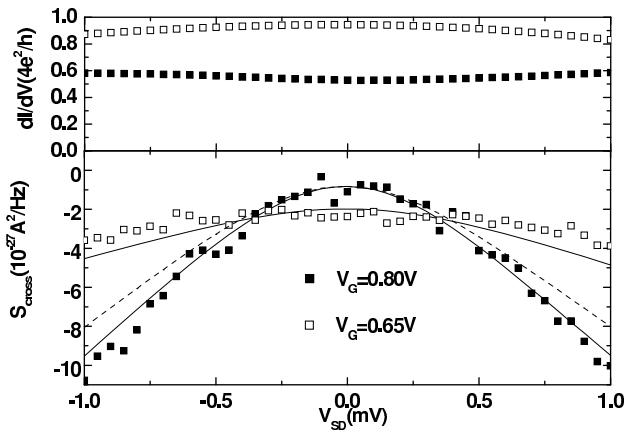}%
\caption{Top: Non-linear conductance as a function $V_{SD}$ for gate
voltages $V_{G}=0.65V$ and $V_{G}=0.80V$. The filled squares
correspond to $V_{G}=0.65V$ and the open squares correspond to
$V_{G}=0.80V$. Bottom: Corresponding noise power spectral density as
a function $V_{SD}$. The lines is formula (\ref{eq:shotnoise}) used
for $T=0.542$ and $T=0.943$ which correspond
to the zero bias value of $dI/dV$ for $V_{G}=0.65V$ and $V_{G}=0.80V$ in units of $4e^2/h$.}%
\label{conductance}%
\end{figure}
The greyscale plot of the non-linear conductance $dI/dV$ as a
function of the gate voltage $V_G$ and the source-drain bias
$V_{SD}$ is displayed on figure \ref{diagramm}a. It exhibits the
characteristic "checker-board" pattern of a Fabry-Perot
interferometer \cite{bockrath:01,Dai:05}. As shown on the side scale
of the greyscale plot, the conductance is modulated from
$0.3\times4e^2/h$ to about $0.95\times4e^2/h$. From the center of
the white "squares" indicated by the dashed lines, one can extract a
value of $3.4meV$ for the level spacing. This value is in good
agreement with the lithographically defined spacing between the $Pd$
electrodes of $500nm$, which yield $hv_F/2L=3.34meV$ for a Fermi
velocity of $8.10^5 m/s$. This value corresponds to the full SWNT
length between the Pd contacts. Therefore, the MWNT contact does not
split the SWNT into two pieces, as previously reported
\cite{GaoBo:04}. The irregularity of the pattern is likely due to
weak scattering. As a consequence, the linear conductance exhibits
sinusoidal oscillations with a changing amplitude, as shown on
figure \ref{frequency}.

The lower panel of figure \ref{conductance} shows the bias
dependence of the current cross-correlations, for gate voltages of
$0.65V$ and $0.80V$, for which the transmission is respectively of
$0.542$ and $0.943$. The data are presented here without any
background correction. For $V_G=0.65V$, the noise power spectral
density starts to display a linear behavior for a bias larger than
$250 \mu V$ which corresponds to $2k_B T\approx 258 \mu eV$ at
$1.5K$. For a lower bias, the noise power spectral density displays
a rounded behavior and saturates. For $V_G=0.80V$, a similar
behavior is observed with a linear regime with a slope
approximatively 3 times smaller than for $V_G=0.65V$.

As shown on figure \ref{conductance}, the conductance is weakly
non-linear in the range of $\pm 1mV$ where the shot noise is
measured. The general formula for the shot noise in a quantum
coherent conductor can in principle account for these
non-linearities \cite{Blanter:00}. Since the maximum variation of
conductance is $10\%$ in the bias range considered, we will assume a
 constant conductance as a function of bias, for the sake of simplicity.
 In this case, if $T_{1,2}$ are the transmissions for the two different orbitals,
the noise reads:

\begin{equation}\label{eq:shotnoise}
S_{I}=\frac{2e^2}{h}\Big( 4k_{B}T\sum_{1,2} T^2_n
+\frac{2eV_{sd}\sum_{1,2}T_n (1-T_n)}{\tanh (\frac{eV_{sd}}{2k_B
T})}\Big)
\end{equation}

If orbital degeneracy is assumed ($T_{1}=T_{2}$), the conductance
completely determines the noise as only the total conductance in
units of $4e^2/h$  enters in equation (\ref{eq:shotnoise}). Figure
\ref{conductance} bottom panel displays in solid curve the shot
noise calculated using the \textit{measured} zero bias total
transmission (top panel) assuming full degeneracy. A quantitative
agreement between the non-interacting theory and the data is found
provided an offset of respectively $6.2\times 10^{-27} A^2/Hz$ and
$10.0 \times 10^{-27}A^2/Hz$ for $V_{G}=0.80V$ and $V_{G}=0.65V$ is
incorporated in formula \ref{eq:shotnoise}.

Combining conductance and shot noise has proved to be an efficient
tool to probe the lifting of spin degeneracy in ballistic conductors
transmitting a single orbital mode \cite{Roche:04}. In the same
spirit, we have investigated a possible lifting of the pseudo-spin
orbital degeneracy in the present nanotube. We find an upper bound
of about $10\%$ for the difference in the transmissions of the two
different orbitals for $V_{G} = 0.65V$. The dashed lines in figure
\ref{conductance} lower panel correspond to the case where
$T_{1}+T_{2}=2 \times 0.542$ but $T_{1}-T_{2}=0.4$, in clear
disagreement with the data.

\begin{figure}[pth]
\centering\includegraphics[height=0.99\linewidth,angle=0]{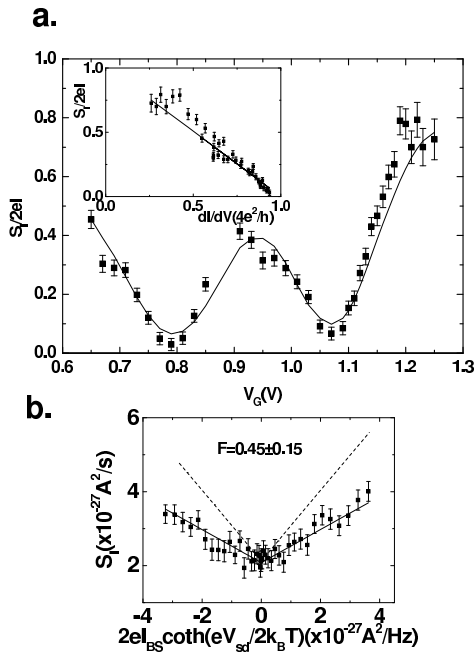}%
\caption{Noise power spectral density measured (filled squares) at
$V_{SD}=-0.7mV$ normalized by the Schottky value $2eI_{SD}$ as a
function of $V_{G}$.
The line
is the theory with the assumption of full orbital degeneracy.
Inset:Noise power spectral density measured (filled squares) at
$V_{SD}=-0.7mV$ normalized by the Schottky value $2eI_{SD}$ as a
function of the measured transmission. The observed linear behavior
is in good agreement with the theory (line). b. Noise power spectral
density as a function of the backscattering current. 
In solid lines , the linear fit gives $F=0.45\pm0.15$. In dashed lines, the two-terminal non-interacting theory ($F=1$)}%
\label{Fano}%
\end{figure}

The transmission dependence of formula \ref{eq:shotnoise} can also
be tested by changing the transmission of the Fabry-Perot
interferometer with the gate voltage. For this purpose, we have
measured the noise for a finite bias voltage $V_{sd}=-0.7mV$
sweeping the gate voltage $V_G$ from $0.65V$ to $1.25V$. In Figure
\ref{Fano}a, the noise power spectral density normalized to the
Schottky value is plotted (filled squares with the error bars) as a
function of the gate voltage which is swept through two resonant
levels. Each point is represented with the statistical error bar
associated to a single averaging run. Note that the shot noise
contribution is obtained here by substraction of background noise,
according to the fitted linear behavior of the inset of figure
\ref{frequency}. The noise displays modulations as a function of the
gate voltage with extrema appearing exactly at the same gate
voltages as for the conductance. Specifically, when the conductance
reaches a maximum, the noise reaches a minimum. As the conductance
maxima at $V_{G}=0.80V$ and $V_{G}=1.05V$ are close to 1 in units of
$4e^2/h$ (respectively 0.943 and 0.90), the noise almost vanishes,
confirming the noiseless character of a fully transmitted fermionic
beam \textit{through a carbon nanotube}. After QPCs \cite{Kumar:96},
carbon nanotubes provide a second example of \textit{noiseless}
conductors. Another remarkable fact is the quantitative agreement of
the measured shot noise with the Fano factor calculated from the
quantum shot noise theory. This is also shown in the inset of figure
\ref{Fano}a where the normalized noise is represented as a function
of the corresponding conductance. As expected, the current noise
displays a linear dependence as a function of the conductance in
units of $4e^2/h$ which vanished for a transmission close to 1.

We now discuss measurements obtained in the weak backscattering
limit which allow in principle a direct determination of the
effective charge transferred through the nanotube. Figure
\ref{Fano}b displays the noise power spectral density for
$V_{G}=0.80V$ as a function of $2eI_{BS}\coth (\frac{eV_{sd}}{2k_B
T})$ (offset substracted) where
$I_{BS}=4e^2/h\int_{0}^{V_{sd}}dV(1-T_{tot}(V))$ is the
backscattering current. A linear slope of $F=0.45\pm0.15$ is
observed (solid lines). In simple cases however, the slope $F$
should be 1 (dashed lines in figure \ref{Fano}b). Interactions in
the Fractional Quantum Hall regime (FQHE) \cite{Saminadayar:97} have
been shown to strongly reduce $F$. However, for the case of a single
mode quantum wire, a similar renormalization as in the FQHE would
imply that the leads are not fermionic
\cite{Ponomarenko:99,Bjorn:02}. Another possibility would be weak
decoherence, possibly induced by the MWNT. Decoherence can indeed be
simulated by adding a third terminal to the circuit
\cite{Blanter:00}. In that case, noise can be lowered with respect
to the pure two terminal (fully coherent) case. Using the
multi-terminal theory of quantum shot noise, we have found that
decoherence could produce a reduction of the shot noise in the weak
backscattering limit. However, it seems difficult to reduce $F$ down
to $0.45$ for the parameters of our sample. We however emphasize
that, since the backscattering current is deduced and not measured
in our experimental setup, we cannot completely rule out a
calibration problem of our setup which would produce such a reduced
shot noise.

In summary, we have measured the zero frequency shot noise of carbon
nanotube based Fabry-Perot interferometers. The noise is modulated
as one sweeps the resonant levels through the Fermi energy of the
reservoirs and vanishes almost as transmission approaches unity. The
data is in quantitative agreement with the non-interacting theory.
In the weak backscattering limit, where shot noise is expected to
follow the Schottky law with the backscattering current, the noise
is found slightly smaller. This may indicate an effect of
electron-electron interactions or weak decoherence.

\begin{acknowledgments}
We thank A. Cottet for a critical reading of the manuscript and N.
Regnault and C. Mora for illuminating discussions. We thank M.
Aprili for giving access to his deposition equipment and P. Hakkonen
and M. B\"uttiker for fruitful discussions. The Laboratoire Pierre
Aigrain (LPA) is the CNRS-ENS UMR8551 associated with universities
Paris 6 and Paris 7. This project is supported by the
ANR-05-NANO-028 contract and by the EU contract FP6-IST-021285-2.
\end{acknowledgments}

\end{document}